# Strain-induced giant enhancement of coercivity in amorphous TbFeCo films

Nattawut Anuniwat[1], Manli Ding[1], S. J. Poon[1] S.A.Wolf[1,2] and Jiwei Lu[2]

[1]*Department of Physics, University of Virginia, Charlottesville, Virginia 22904, USA*

[2]*Department of Materials Science and Engineering, University of Virginia, Charlottesville, Virginia 22904, USA*

[a)] Corresponding author: jl5tk@virginia.edu, na8x@virginia.edu

**Abstract:**

We report a strong size dependence of coercivity in amorphous ferromagnetic TbFeCo films. The as-deposited film exhibited a low saturation magnetization ($M_S$ 100 emu/cc) and a high perpendicular anisotropy ($K_U$ $10^6$ erg/cc). Hall-bar devices were fabricated for characterizing the magneto-transport behaviors. A significant increase in coercivity (up to ~ 300 %) was observed at room temperature as the width of Hall bar was reduced. The large coercivity enhancement was attributed to the relaxation of film stress. The effect of strain and dimensionality on the coercivity in TbFeCo makes it attractive for tunable coercivity and the magnetization reversal in future nanoscale devices.

## I. INTRODUCTION

The ferrimagnetism in amorphous rare-earth (RE) transition metal (TM) alloys is well known, and has been investigated for the magneto-optical recording due to perpendicular magnetic anisotropy (PMA) in thin films[1]. Recently, RE-TM alloys have been investigated for applications in perpendicular magnetic random access memory (p-MRAM), which is considered to be a universal memory technology due to the low power dissipation and the non-volatility[2]. It has been reported that amorphous TbFeCo was used to form perpendicular magnetic tunnel junctions (p-MTJ) for p-MRAM application[3].

The origin of the perpendicular anisotropy in RE-TM alloys remains elusive. Studies have correlated the magnetic anisotropy of RE-TM magnetic films with various structural characteristics ranging from columnar textures[4] to microcrystallinity[5] to local magnetic or/and local structural anisotropy[6,7]. The coupling between RE and TM ions is antiferromagnetic[6]. The large TM-TM ferromagnetic interaction essentially align the TM moments, while the large RE anisotropy fan the RE moments out over the opposite hemisphere as in the Dy-Fe system. This results in a low magnetization for RE-TM alloy, and likely the large magnetorestriction effect, as reported in DyFeCo[8]. Nonetheless, the magneto-transport behavior in RE-TM alloy particularly TbFeCo has not yet been systematically explored up to date.

We have obtained amorphous TbFeCo (Tb content ~ 30 at. %) films via the combinatorial growth technique. The specific compositions were chosen due to high coercive field which lead to high thermal stability, making the materials suitable for device application. We report the magneto-transport behavior of TbFeCo with a high perpendicular anisotropy. A significant change in the coercive field was observed as a function of the width of TbFeCo Hall bars. The possible mechanisms for this coercivity enhancement will be discussed.

## II. EXPERIMENT

TbFeCo films were grown on thermally oxidized Si substrates by *rf* magnetron sputtering at ambient temperature. Tb, Fe, and Co targets were used concurrently during the sputtering so that a window of composition was explored to determine the optimal composition for the PMA. The base pressure of the vacuum chamber was ~ $5\times10^{-7}$ Torr. The deposition was performed in ultra-high purity Ar gas under the pressure of 5 mTorr. For the TbFeCo films studied in this letter, the powers applied to Tb, Fe and Co targets were 40 Watts, 89 Watts and 35 Watts respectively, which yielded the growth rate of ~ 5 nm/minute. Subsequently a ~ 10 nm MgO capping layer was sputtered to protect the surface of TbFeCo film.

The composition of $Tb_{30}Fe_{63.5}Co_{6.5}$ was determined by X-ray photoelectron spectroscopy. The microstructure was characterized by X-ray diffraction (XRD), and the thickness of the film was measured by X-ray reflectivity (Smart-lab®, Rigaku Inc.). Transmission electron microscopy (TEM) was also used to examine the microstructure. TEM cross-section samples were prepared by standard TEM sample preparation techniques, with Ar ion milling as the final step. A Titan (FEI) transmission electron microscope was used. Atomic Force Microscopy and Magnetic Force Microscopy (Cypher™, Asylum Research Inc.) were used to characterize surface morphology and magnetic domain structure. Windows Scaning Microscope (WSxM) software was used to determine the size of magnetic domain.

Both in-plane and out-of-plane hysteresis loops were measured using vibrating sample magnetometer (VSM) and magneto-optical Kerr Effect (MOKE) looper. Hall bar devices were fabricated to characterize the magneto-transport behaviors, by using the photolithography process and the wet etching in the diluted hydrochloride acid. The width of Hall bars ranged from 50 – 500 μm. The top contact was 200 nm Au/ 20 nm Ti deposited by e-beam evaporation. The Hall resistance and magneto-resistance were measured in a Versa Lab system (Quantum Design Inc.)

**III. RESULTS AND DISCUSSION**

Fig. 1 shows TEM images of a ~35 nm thick TbFeCo films. No evidence for crystalline structures was observed in the cross-section TEM bright field contrast image as shown in Fig. 1a. High-resolution TEM image (Fig. 1b) showed no nano-crystallites inside the film. Only a single ring pattern was presented in the fast Fourier transform (FFT) image that attributed to the lack of long range ordering. In addition, XRD scans showed that there were no diffraction peaks other than these from the substrate for TbFeCo films with different thickness (not shown here). Both TEM and XRD confirmed the amorphous nature of the film deposited at the ambient temperature, which was consistent with the previous reports on RE-TM films[4,5].

Fig. 2a shows hysteresis loops of a 35 nm thick TbFeCo film. The saturation moment ($M_s$) was ~ 100 emu/cc. The small saturation moment was due to the ferrimagnetism of amorphous RE-TM alloys. Co and Fe ions have a small magnetic anisotropy and Tb ions have large anisotropy. The large TM-TM ferromagnetic interaction aligns the magnetic moments among Fe and Co ions, while the large anisotropy of RE ions fans the magnetic moments of Tb out over the opposite hemisphere. As a result, the net moment was the difference of magnetic moment between Tb and Co/Fe. The out-of-plane coercive field ($H_c$) of this sample was ~ 0.66 T and magnetic anisotropy field ($H_K$) was ~ 1.6 T. From the hysteresis loops one can estimate the anisotropy energy ($K_U$) ~ 0.8 x $10^6$ erg/cc that was out of the plane.

Figure 2b shows the Hall Resistance ($R_H$) as a function of an external magnetic field, which was applied perpendicular to the film plane, for 50 and 500 μm wide Hall bars at room temperature. There was a significant difference in $H_c$ for two different Hall bars measured via this method. The magnetic field dependence of Hall resistances for both Hall bars resembled the hysteresis loop of out-of-plane magnetization of the samples, which was due to the dominance of the anomalous Hall resistance ($R_{AHE}$) over the ordinary Hall resistance ($R_{OHE}$). $R_{AHE}$ was proportional to the magnetization ($M$) of the ferromagnet, thus it was strongly dependent to the magnetic field. Based on the relationship between $R_{AHE}$ and $M$, the values of $H_c$ of Hall bars were extracted as a function of the width of Hall bar. For the 500 μm wide Hall bar, $H_c$ was 0.6 T that was close to the value (~0.66 T) measured from out-of-plane hysteresis loop that was consistent with the value measured on unpatterned film by VSM and MOKE. In contrast, $H_c$ was increased to ~2.1 T when the width of the Hall bar was reduced to 50 μm, representing a three fold increase over $H_c$ measured from both 500 μm wide Hall bar and the unpatterned film.

Similar size dependence was also observed in the magneto-resistances of these two Hall bars, as shown in Fig 2c. Two sharp peaks were observed at the magnetic fields corresponding to $H_c$ in AHE loops shown in Fig 2b. As the magnetic field increase in the positive direction, the resistance remained almost constant up to the positive $H_c$ where resistance suddenly increase creating a positive sharp peak, then the resistance decreased to almost the same value as before and remained constant value through the maximum 3

T. The same MR behavior was also observed as the magnetic field decrease. There was also a sharp peak at negative $H_c$. However, this peak was negative.

These two antisymmetric magneto-resistance peaks were also observed in Pt/Co multilayer structures with PMA. MR peaks are strongly related with the domain number per unit area, and occur near the coercive field as a result of multi-domain state of the film during the magnetization reversal process. The opposite magnetizations across the domain wall cause the opposite Hall fields. This induced the circulating current around domain wall, leading to the redistribution of MR peak and electric field[9]. The antisymmetric MR indicates that domain wall direction is in the film plane and is perpendicular to the current direction. Nonetheless, the occurrence of MR peaks at $H_C$ confirmed the strong size dependence of coercivity in amorphous TbFeCo. On the other hand, the MR ratios (defined as $(R(H)-R(0))/R(0)\times 100\%$) for 500 μm and 50 μm were similar, 1.4 % and 1.6 % respectively, which was likely due to the similar densities of magnetic domains as previous reported[10].

Additional study on the effect of the film thickness was also performed. Fig 2d shows $H_C$ as a function of Hall bar width measured at room temperature while the thickness of TbFeCo was changed from 15 nm to 100 nm. The unpattern films coercivity increased from ~0.2 T for a 15 nm thick TbFeCo film to 1 T for a 100 nm TbFeCo film monotonically. The coercivity reached a plateau showing a very small change as the thickness of the film increased from 60 nm to 100 nm. The effect of film thickness on the enhancement of $H_C$ in the TbFeCo Hall bar was studied. We observed the similar enhancement in coercive field as width of the Hall Bar is reduced except for the 15 nm and 20 nm thick films, in which $H_C$ remained the same as unpatterned films.

Figure 3 shows the MFM images taken from 50 μm wide Hall bars with the different film thicknesses. The MFM were performed after demagnetizing. There were little topographic differences among the devices based on the AFM images taken with MFM images (not shown here). The surfaces of films appeared to be uniform. The root-mean-square roughness was approximately 3 nm for all samples. The contrast in MFM images showed the strip domain structure that was typical for magnetic thin films with the perpendicular magnetization anisotropy. The domain size and shape became less uniform in the 15 nm and 30 nm thick films as compared to the strip domains in the films with the thickness over 30 nm.

The size of magnetic domain versus the thickness of films was summarized in Table I. The size of magnetic domain was increased monotonically with film thickness. The average distance between nearest neighbor domain was extracted to represent the density of the magnetic domains in a given area, in this case 100 μm$^2$. The domain width was proportional to $(tK_U)^{1/2}$ where $K_U$ is perpendicular anisotropy and $t$ is film thickness[11].

The coercive field is usually associated with the magnetization reversal process, and various effects on the reversal process can be linked to the dimensionality of magnetic wires, particularly in crystalline L1$_0$ FePt as recently reported[12,13]. It has been reported that the coercivity of FePt nanowires was increased in narrow Hall bars with the width less than 100 nm thanks to the changes in the nucleation and propagation of magnetic domains[12]. In contrast, the coercivity in Hall bars with the width over 200 nm, made from L1$_0$ FePt, was increased by the pinning field of domain wall from the defects[13]. However, the enhancement in TbFeCo Hall bars cannot be attributed to domain structure or lattice defects. Firstly, the domain size (0.1~ 0.2 μm$^2$) was too small in comparison with the width of Hall bars (50~100 μm) to play an important role in the change in the magnetism reversal process or the magnetic anisotropy. Also the magnetic domain structure appeared to be very similar regardless the width of Hall bars. Secondly, the absence of long rang ordering in TbFeCo films made it less susceptible to the pinning fields associated with the defects.

The effect of the edge roughness on the coercivity was also considered. The additional strength of the magnetic field was calculated followed the equation given by Cayssol *et al.*[14] The estimated domain wall energy $\sigma$ was ~ $10^{-3}$ J/m$^2$, with $M_s$ of ~ $10^5$ A/m$^3$ and the roughness < 0.002 (from SEM observation). The extra coercive field required to overcome the rough edge was less than $10^{-6}$ Tesla, which was much smaller than the extra coercivity that has been observed.

In the absence of magneto-crystalline anisotropy, the enhancement of coercivity may be explained by the stress-induced anisotropy in amorphous TbFeCo films due to the relaxation of compressive strain in Hall bar devices thanks to the large magnetostriction effect. The magnetostriction effect on the perpendicular magnetic anisotropy has been shown for various magnetic thin films[8,15]. In particular, Saito et al. have shown that the tensile strain in DyFeCo alloy modulated the easy axis of magnetization[8]. Amorphous TbFeCo films experienced the growth stress resulted from incorporation of excess atomic density due to the surface stress. The film strain in the film was bi-axial and was partially relaxed in the Hall bar structures due to the reduction in the surface stress. The relaxation of film strain resulted in the increase in the coercivity of TbFeCo films. The surface stress was reduced further with the width of Hall bar, giving rise to the width dependence of the coercivity in the films thicker than 30 nm. In comparison, we did not observe any effect of the strain relaxation in the 10 and 20 nm films probably because the width of Hall bars was not small enough to release the surface stress significantly. One can expect the enhancement in the coercivity when the width of Hall bars is further reduced. To further quantify the magnetostriction effect in TbFeCo, one needs to exert the mechanical strain through a bending apparatus and monitor the magnetization reversal process through the anomalous Hall effect.

## IV. CONCLUSIONS

We have synthesized amorphous TbFeCo with ~ 30 at. % Tb films via the combinatorial growth, and reported the magneto-transport behavior of TbFeCo with perpendicular anisotropy using anomalous Hall effect, and observed a significant increase in coercive field as a function of the width of TbFeCo Hall bars. The enhancement in coercivity was likely linked to the relaxation of compressive strain due to the reduction in the surface stress. The effect of strain and dimensionality on the coercivity in TbFeCo will make it an attractive candidate for future nanomagnetics and spintronics not only for the higher thermal stability (sufficient magnetic anisotropy) in nanoscale devices, but also for the tunable coercivity and the magnetization reversal via the strain or stress (the magnetostriction effect).

**ACKNOWLEDGEMENT**

Authors are grateful to the financial support from DTRA (Award number: HDTRA 1-11-1-0024) and DARPA (Award number: HR0011-09-C-0023).

**Reference:**

1. S. Tsunashima, Journal of Physics D-Applied Physics **34** (17), R87 (2001).

2. E. Chen, D. Apalkov, Z. Diao, A. Driskill-Smith, D. Druist, D. Lottis, V. Nikitin, X. Tang, S. Watts, S. Wang, S. A. Wolf, A. W. Ghosh, J. W. Lu, S. J. Poon, M. Stan, W. H. Butler, S. Gupta, C. K. A. Mewes, T. Mewes, and P. B. Visscher, Ieee Transactions on Magnetics **46** (6), 1873 (2010).

3. M. Nakayama, T. Kai, N. Shimomura, M. Amano, E. Kitagawa, T. Nagase, M. Yoshikawa, T. Kishi, S. Ikegawa, and H. Yoda, Journal of Applied Physics **103** (7) (2008); C. M. Lee, L. X. Ye, T. H. Hsieh, C. Y. Huang, and T. H. Wu, Journal of Applied Physics **107** (9) (2010).

4. H. J. Leamy and A. G. Dirks, Journal of Applied Physics **50** (4), 2871 (1979).

5. Y. Takeno, K. Kaneko, and K. Goto, Japanese Journal of Applied Physics Part 1-Regular Papers Short Notes & Review Papers **30** (8), 1701 (1991).

6. J. M. D. Coey, J. Chappert, J. P. Rebouillat, and T. S. Wang, Physical Review Letters **36** (17), 1061 (1976).

7. V. G. Harris, K. D. Aylesworth, B. N. Das, W. T. Elam, and N. C. Koon, Physical Review Letters **69** (13), 1939 (1992).

8. N. Saito, M. Yamada, and S. Nakagawa, Journal of Applied Physics **103** (7) (2008).

9. X. M. Cheng, S. Urazhdin, O. Tchernyshyov, C. L. Chien, V. I. Nikitenko, A. J. Shapiro, and R. D. Shull, Physical Review Letters **94** (1) (2005).

10. S. T. Li, T. Amagai, X. X. Liu, and A. Morisako, Applied Physics Letters **99** (12) (2011).

[11] Z. Liu, S. M. Zhou, and X. B. Jiao, Journal of Physics D-Applied Physics **42** (1) (2009).

[12] V. D. Nguyen, L. Vila, A. Marty, J. C. Pillet, L. Notin, C. Beigne, S. Pizzini, and J. P. Attane, Applied Physics Letters **100** (25) (2012).

[13] J. P. Attane, D. Ravelosona, A. Marty, V. D. Nguyen, and L. Vila, Physical Review B **84** (14) (2011).

[14] F. Cayssol, D. Ravelosona, C. Chappert, J. Ferre, and J. P. Jamet, Physical Review Letters **92** (10) (2004).

[15] A. Bur, T. Wu, J. Hockel, C. J. Hsu, H. K. D. Kim, T. K. Chung, K. Wong, K. L. Wang, and G. P. Carman, Journal of Applied Physics **109** (12) (2011).

TABLE I. Comparison of domain structure in 50 μm wide Hall bars.

| Thickness(nm) | Domain Size (μm$^2$) | S$^*$ (μm) |
| --- | --- | --- |
| 15 | 0.103 | 0.147 |
| 30 | 0.112 | 0.165 |
| 60 | 0.224 | 0.305 |
| 100 | 0.183 | 0.207 |

$^*$S is average distance between two nearest neighboring domains.

**Figure captions:**

**Figure 1:** (a) Bright field TEM image of a 35 nm TbFeCo film deposited on $SiO_2$/Si. A 15 nm MgO layer was used to cap the film; (b) HRTEM image of the TbFeCo film. The inset is a FFT pattern of the image.

**Figure2:** (a) Hysteresis loops of a 35 nm TbFeCo film measured at room temperature; (b) Hall resistance vs. magnetic field for two Hall bar devices with the widths of 50 and 500 μm. The inset is an optical image of a Hall bar device; (c) Magneto resistance vs. magnetic field for two Hall bar devices with the widths of 50 and 500 μm; (d) Coercive field vs. the width of Hall bar for TbFeCo films with various thicknesses measured at room temperature

**Figure 3:** Magnetic force microscopy images of Hall bar devices with the width of 50 μm with various film thickness (a) 15 nm (b) 30 nm (c) 60 nm (d) 100 nm.

**Figure 1**

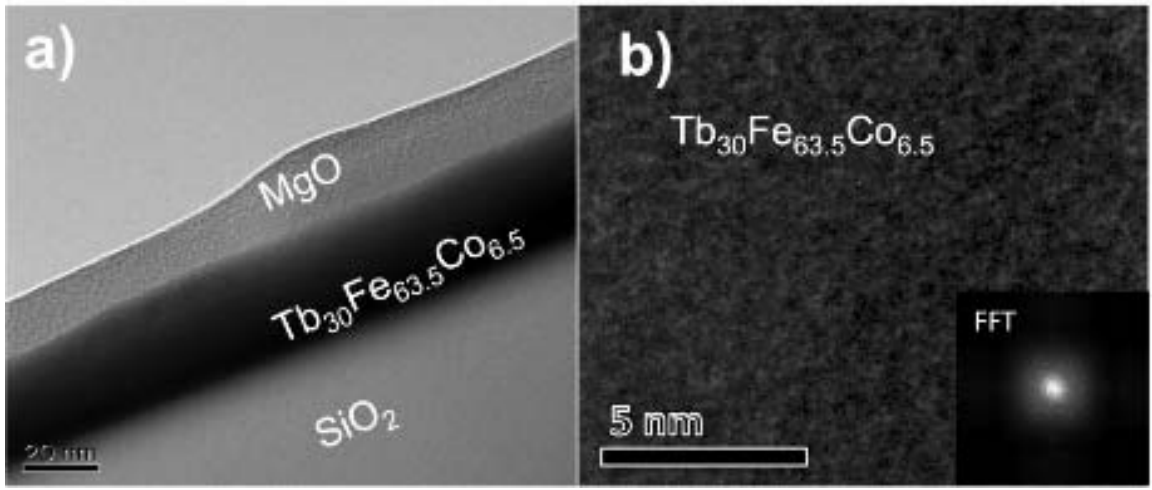

**Figure 2**

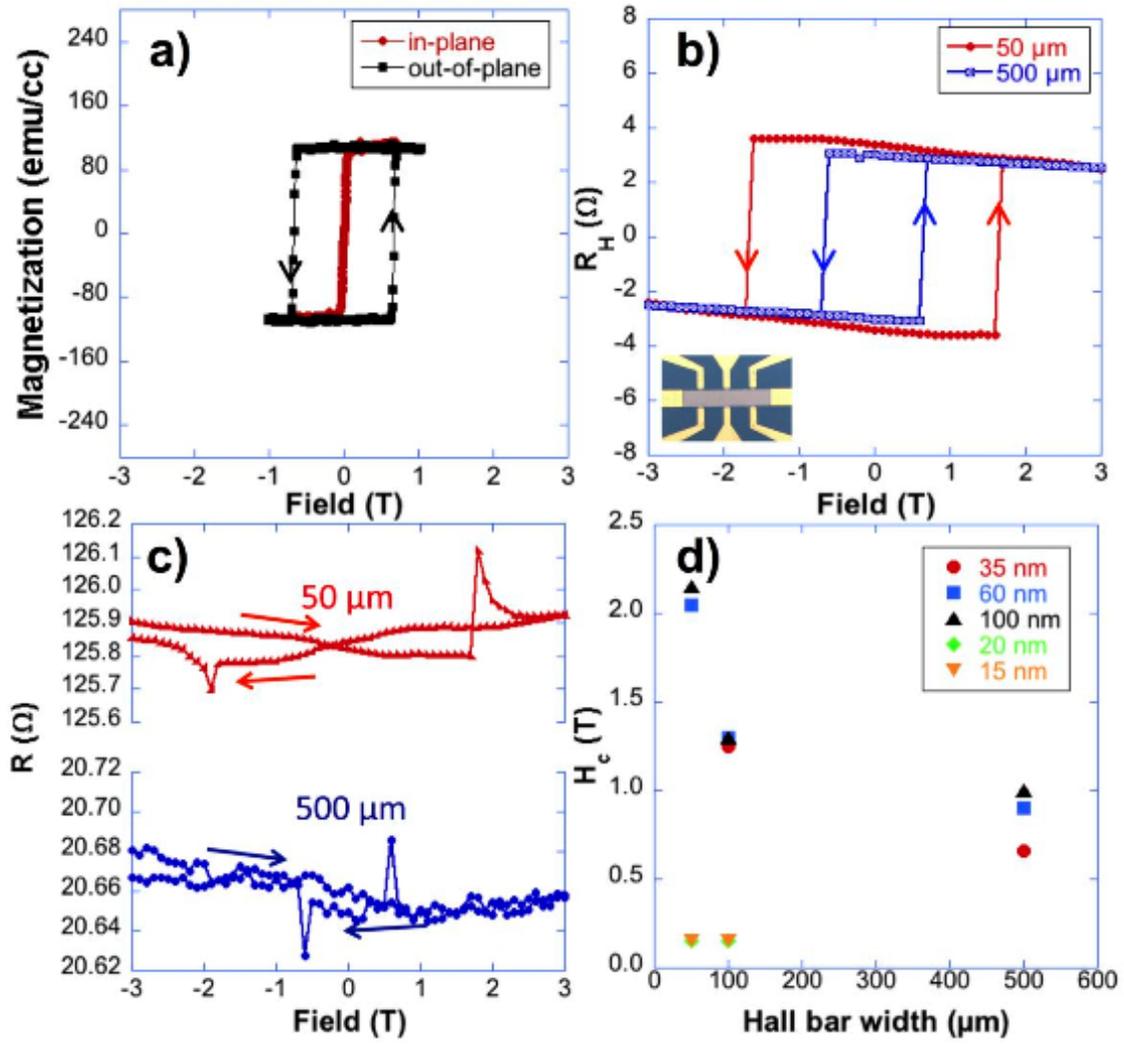

**Figure 3**

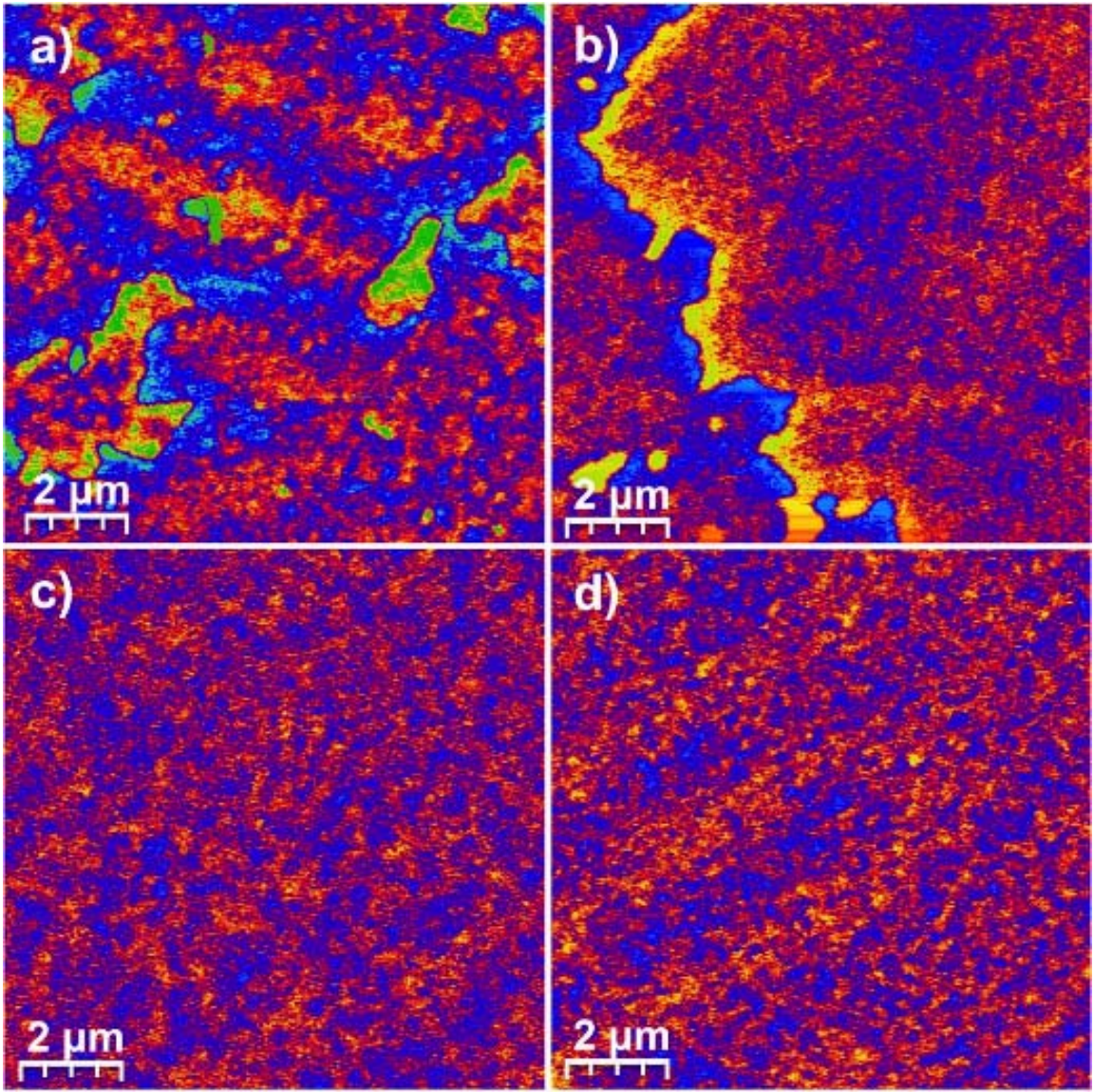